\documentstyle[epsf,prl,aps]{revtex}
\begin{document}
\twocolumn[\hsize\textwidth\columnwidth\hsize\csname@twocolumnfalse\endcsname
\preprint{\vbox{\hbox{November 1997} \hbox{IFP-746-UNC}  }}
\title{3-3-1 exotic quark search at CERN LEPII-LHC}
\author{Y. A. Coutinho}
\address{Instituto de F\'\i sica, Universidade Federal do Rio de Janeiro, Ilha do Fund\~ao, 21945-970 Rio de Janeiro, RJ, Brazil}
\author{P. P. Queir\'oz Filho and M. D. Tonasse}
\address{Instituto de F\'\i sica, Universidade do Estado do Rio de Janeiro, Rua 
S\~ao Francisco Xavier, 524, 20550-013 Rio de Janeiro, RJ, Brazil}
\maketitle

\begin{abstract}
The 3-3-1 electroweak model is the simplest chiral extension of the standard 
model which predicts single and double charged bileptons and exotic quarks 
carrying -4/3 and 5/3 units of the positron charge. In this paper we study the 
possibilities of the production and decay of one of these exotic quarks at CERN 
LEPII-LHC collider. For typical vector bilepton, exotic quark masses and mixing 
angles we obtained between 20 and 750 events per year. Angular distributions are 
also presented.
\end{abstract}
\vskip1pc]
\bigskip
\bigskip

\narrowtext
\section{Introduction}
\label{intro}
It is a celebrated fact that the standard model of the electroweak interactions 
accommodates all the present experimental results. However, since it is not able 
to give response to some fundamental questions in particle physics, the building 
of extensions and alternative models are well motivated. One of these questions 
is the family replication problem, which can have an elegant solution in the 
simplest chiral extension of the standard model \cite{PP92,FR92}. We are 
referring on a chiral model which is based on the 
SU(3)$_C\otimes$SU(3)$_L\otimes$U(1)$_N$ (3-3-1 for short) semi simple symmetry 
group which breakdown to the SU(3)$_C\otimes$SU(2)$_L\otimes$U(1)$_Y$  in some 
energy scale higher than the Fermi one.\par
One peculiar feature of the model is that the anomaly cancelation occurs only 
when the three fermion generations are considered together. This implies that 
the number of families must be multiple of the color number and, as a 
consequence, the 3-3-1 model suggests a route towards the response of the flavor 
question.\par
There is a high interest in the bilepton phenomenology (see Ref. \cite{CD98} and 
references cited therein). Usually bileptons ($L= 2$) are vector gauge or scalar 
bosons which couple two leptons. In 3-3-1 model they couple also an ordinary to 
an exotic quark. Some works on 3-3-1 phenomenomenology can be found in Refs. 
\cite{DJ99,PS98,JJ97,AF96,ST95,FL94,LN94,DP94,NG94,FR94}. The presence of the 
vector bileptons in the model has two phenomenologically interesting 
possibilities. First there is a hope that this kind of gauge bosons can be 
detected in a relatively low mass scale and second they contribute to processes 
which violate lepton number, {\it i.e.}, it is free of standard model 
background. Since a new generation of colliders will be working in the next 
years, we think that this kind of model deserves more detailed phenomenological 
treatment.\par
In this paper we are interested in 3-3-1 exotic quark production and decay, 
particularly one carrying 5/3 units of positron charge. This exotic quark, if 
detected, would be also a signature for a double charged bilepton present in the 
model. Double charged vector bilepton masses are bounded from below from muonium 
to antimuonium conversion to a value $\sim$ 850 GeV \cite{Wet99}.\par
Here a comment is in order. All the constraints on the 3-3-1 parameters coming 
from experiments evolving leptonic interaction should be seen with care. In 
3-3-1 model the leptons mix by a Cabibbo-Kobayashi-Maskawa like mixing matrix 
whose elements do not yet measured \cite{LN94}. Usually these experiments apply 
only when the leptonic mixing matrix is diagonal \cite{Wet99}. Also, in models 
with extended Higgs sector some not unrealist situations could exist in which 
scalar bosons contribution to muonium to antimuonium conversion is not 
negligible \cite{HW96}. Therefore, in this work we assume a more 
interesting lower limit on the double charged bilepton mass, $\sim 350$ GeV, 
which is accessible to next generation of accelerators and is compatible with 
others low energy bounds \cite{FL94,FN92}.\par
Recently, lower bounds on exotic supersymmetric particles were translated to a 
lower bound on 3-3-1 exotic quark masses $\sim$ 250 GeV \cite{DJ99}. Upper bound 
on bilepton masses can be $\sim$ 3.5 TeV \cite{JJ97} and the exotic quark masses 
have no upper bound.\par
We examine here the possibility for the observation of the reaction $e^- p 
\longrightarrow e^+ X l^- l^-$ at CERN LEPII-LHC center of mass energy, where 
$X$ is a quark jet and $l = e, \mu, \tau$. The 3-3-1 exotic quark production 
{\it via} double charged bilepton exchange was already studied \cite{ST95}. 
However, our results differ from the previous one since we considered also the 
decay of the exotic quark and mixing angles.\par 
This paper is organized as follows. In Sec. \ref{sec1} we discuss the relevant 
features of the 3-3-1 model, in Sec. \ref{sec2} we present the total cross 
sections and some distributions and finally in Sec. \ref{sec3} our conclusions.

\section{The 3-3-1 Model}
\label{sec1}
Let us summarize the most relevant points of the model. In its minimal version 
the fermion representation content is
\begin{mathletters}\begin{eqnarray}
\psi_{aL} & = & \left(\begin{array}{c}\nu_a \\ l^\prime_a \\ {l^\prime}^c_a
\end{array}\right)_L \sim \left({\bf 3}, 0\right);\\
Q_{1L} & = & \left(\begin{array}{c} u^\prime_1 \\ d^\prime_1 \\ J_1
\end{array}\right)_L \sim \left({\bf 3}, \frac{2}{3}\right), \\
Q_{\alpha L} & = & \left(\begin{array}{c} J^\prime_\alpha \\ u^\prime_\alpha \\
d^\prime_\alpha
\end{array}\right)_L \sim \left({\bf 3}^*, -\frac{1}{3}\right),
\end{eqnarray}\label{quark}\end{mathletters}
where $l^\prime_a = e^\prime, \mu^\prime, \tau^\prime$, $\alpha$ = 2, 3 
\cite{PP92}. The primed fields are symmetry eigenstates. In Eqs. (\ref{quark}) 
0, 2/3 and $-$1/3 are the U(1)$_N$ charges. In this work we are considering 
massless neutrinos \cite{FR94}. Each left-handed quark field has its 
right-handed counterpart transforming as a singlet of the SU(3)$_L$ group. In 
order to avoid anomalies one of the quark families must transform in a different 
way with respect to the two others. In fact, the first and the third generation 
of quarks were arbitrarily singularized by the authors of the Refs. \cite{PP92} 
and \cite{FR92}, respectively, but can be showed that we can map one 
representation to another representation performing an unitary transformation 
\cite{DP94,NG94}. The $J_1$ exotic quark carries 5/3 units of electric charge 
while $J_2$ and $J_3$ carry $-$4/3 each. The exotic quarks couple to the 
ordinary ones {\it via} bileptons which leads to processes where the total 
lepton number conservation is violated. Should be notice that in the leptonic 
sector of the model the particle spectrum coincides with the standard model 
one.\par
In the gauge sector the single charged $\left(V^\pm\right)$ and the double 
charged $\left(U^{\pm\pm}\right)$ vector bileptons, together with a new neutral 
gauge boson $Z^{0\prime}$ complete the particle spectrum with the charged 
$W^\pm$ and the neutral $Z^0$ standard gauge bosons.\par
The charged current interactions for the quarks are given by 
\begin{eqnarray}
{\cal L}_q & = & -\frac{g}{2\sqrt{2}}\left[\overline{U}\gamma^\mu(1 - 
\gamma_5)V_{\rm CKM}DW^+_\mu + \right.\cr
&&\left. + \overline{U}\gamma^\mu(1 - \gamma_5)\zeta{\cal 
JV}_\mu + \overline{D}\gamma^\mu(1 - \gamma_5)\xi{\cal JU}_\mu\right] + \cr 
&& + {\mbox 
{\rm H. c.}},
\label{lq}
\end{eqnarray}
where we modified slightly the notation of the reference\cite{PP92}, with
\begin{mathletters}\begin{eqnarray}
U = \left(\begin{array}{c} u \\ c \\ t
\end{array}\right), \quad
D & = & \left(\begin{array}{c}
         d \\ s \\ b
\end{array}\right), \quad
{\cal V}_\mu = \left(\begin{array}{c}
         V^+_\mu \\ U^{--}_\mu \\ U^{--}_\mu\end{array}\right),\\
{\cal U}_\mu & = & \left(\begin{array}{c}
         U^{--}_\mu \\ V^+_\mu \\ V^+_\mu
\end{array}\right), 
\end{eqnarray}\label{maest}\end{mathletters}
and ${\cal J} = {\rm diag}\left(\begin{array}{ccc}J_1 & J_2 & 
J_3\end{array}\right)$. 
The $V_{\rm CKM}$ is the usual Cabibbo-Kobayashi-Maskawa mixing matrix and $\xi$ 
and $\zeta$ are mixing matrices containing new unknown mixing parameters due to 
the presence of the exotic quarks. In the leptonic sector we have the charged 
currents
\begin{eqnarray}
{\cal L}_l & = & -\frac{g}{2\sqrt 
2}\sum_l\left[\overline{\nu}_l\gamma^\mu\left(1 - 
\gamma_5\right)lW^+_\mu +\right.\cr
&&\left. + \overline{l}^c\gamma^\mu\left(1 - 
\gamma_5\right)\Lambda\nu_lV^+_\mu +\right.\cr
 &&\left.+ \overline{l}^c\gamma^\mu\left(1 - 
\gamma_5\right)\Delta lU^{++}_\mu\right] + \mbox{H. c.},
\label{ll}
\end{eqnarray}
where $\Lambda$ and $\Delta$ are lepton mixing matrices. In Eqs. (\ref{lq}), 
(\ref{maest}) and (\ref{ll}), unlike Eq. (\ref{quark}), we are working with the 
mass eigenstates. There are not measurements for matrix elements of $\xi$, 
$\zeta$, $\Lambda$ and $\Delta$ mixing matrices. An estimative for $\xi^*_{J_1d} 
\xi_{J_1s} $ was obtained from the neutral kaon system mass difference 
\cite{PS98}.\par
The pattern of symmetry breaking is SU(3)$_L\otimes$
U(1)$_N$${\longmapsto}
$SU(2)$_L\otimes$U(1)$_Y$${\longmapsto}$U(1)$_{\rm em}$. The bileptons and 
exotic quarks get masses when the 3-3-1 symmetry breakdown to the standard model 
one. This breakdown determines the scale for the new physics in the model. At 
low energy the 3-3-1 model recovers the standard phenomenology\cite{NG94}.

\section{Cross Sections and Distributions}
\label{sec2}
\begin{figure}[h]

\begin{center}

\epsfxsize=2.8in

\epsfysize=10 true cm

\ \epsfbox{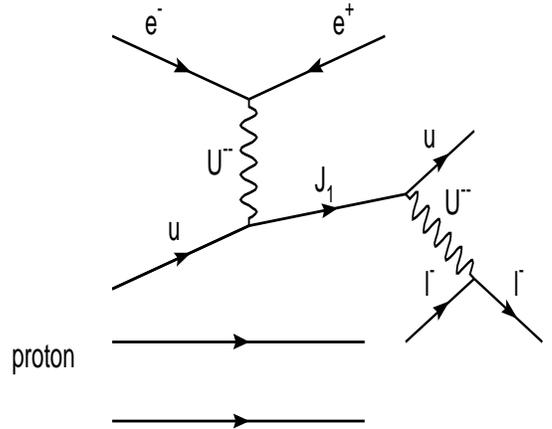}

\end{center}
\vskip -4 true cm

\caption[]{Feynman diagram for the process $e^- p$ $\longrightarrow$ $e^+ l^- 
l^- X$.}

\label{fig:1}

\end{figure}

In this work we studied the production and decay of exotic quarks from 3-3-1 
model in $e^-p$ collision for CERN LEPII-LHC center of mass energy through the 
deep inelastic scattering (DIS), represented in diagram of the Fig. \ref{fig:1}.\par

The production and decay of heavy exotic quarks at HERA energy was studied in a 
different context by one of the authors. In that paper was investigated the 
signatures for heavy quarks using vector singlet, vector doublet and 
fermion-mirror-fermion extended models \cite{AS94}. Here, we study the 
production and decay of $J_1$ quark  {\it via} double charged bilepton exchange. 
We are motivated by the following particular features of the model: (a) the 
absence of standard model background to $e^- p \longrightarrow l^+ X l^-l^-$ 
reaction (Fig. \ref{fig:1}), due to leptonic number violation; (b) the possible low mass 
scale of bileptons which contributes to give reasonable rate of events in the 
next generation of colliders \cite{CD98}; (c) the production of a charged 
primary lepton (positron) that can be detected and (d) the existence of quarks 
with fractionary charge greater than one.\par
\begin{figure}[h]

\begin{center}

\epsfxsize=3.8in

\epsfysize=9 true cm

\vskip -1 true cm \hskip -1.5 true cm
\ \epsfbox{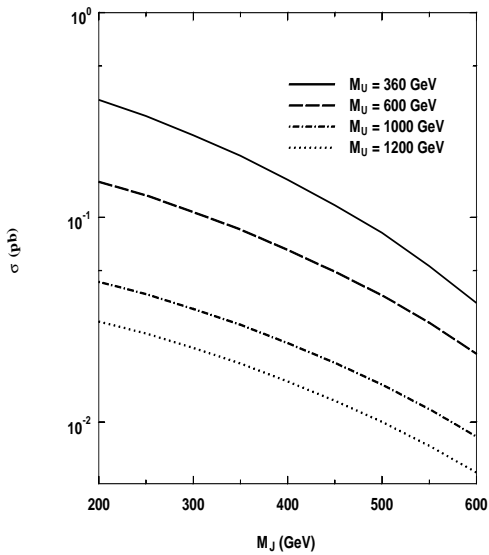}

\end{center}

\vskip -0.5 true cm
\caption[]{Cross section as a function of $M_J$ for various values $M_U$.}

\label{fig:2}

\end{figure}

The $e^-p$ collision is an interesting place for 3-3-1 model phenomenological 
study, because presents an unusual vertex were a 3-3-1 bilepton couples an 
ordinary with an exotic  quark [as can see in Eq. (\ref {lq})].\par
\begin{figure}[h]

\begin{center}

\epsfxsize=3.8in

\epsfysize=9 true cm

\vskip -1.8 true cm \hskip 1 true cm
\ \epsfbox{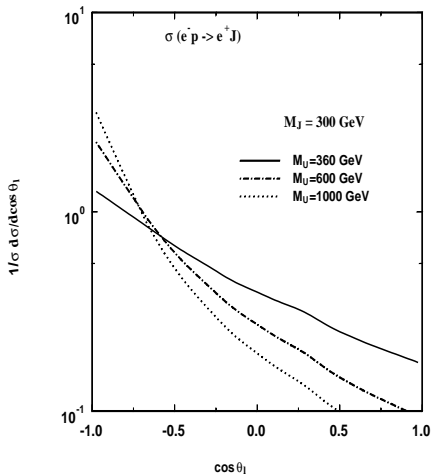}

\end{center}

\vskip -1 true cm
\caption[]{Normalized angular distribution of the primary lepton with $M_J
= 300$ GeV for various $M_U$.}
 
\label{fig:3}

\end{figure}

An exotic $J$ quark can be produced through DIS and vector-boson-gluon-fusion 
(BGF). The BGF process is suppressed by the mixing parameters when compared with 
standard model top quark production. 
In the lack of a more realistic estimate, we can suppose that this mixing
parameter behaves as the one estimate in Ref. \cite{PS98}. There, under
reasonable hypothesis, was showed that numerically the upper bound of the
mixing parameter contributing to $K^0-\bar K^0$ transition  depends linearly
on the bilepton mass as  $\xi^*_{J_1d} \xi_{J_1s} \simeq 0.1  M_U$, when $M_U$
is the double charged bilepton mass in TeV.\par
\begin{figure}[h] 
\begin{center} 
\epsfxsize=3.8in 
\epsfysize=9 true cm 
\vskip -1.8 true cm \hskip 1 true cm
\ \epsfbox{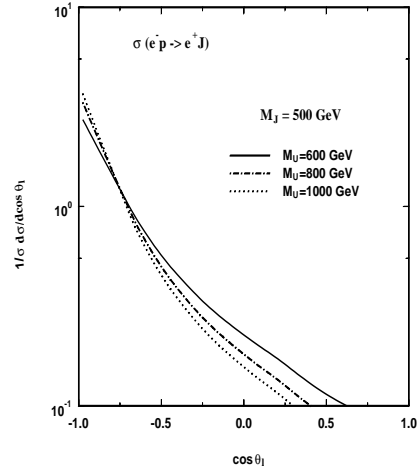} 
\end{center} 
\vskip -1 true cm
\caption[]{The same as Fig. \ref{fig:3} but for $M_J =
500$ GeV.}  \label{fig:4} 
\end{figure}
However, since this  parameter
appears in the cross section formula as a multiplicative factor, our  results
can be immediately adapted for other values of the mixing angles (should  be
notice that this observation apply also to the leptonic mixing parameters). 
\begin{figure}[h]
 
\begin{center}

\epsfxsize=3.8in

\epsfysize=9 true cm

\vskip -1.6 true cm 
\hskip 1 true cm
\ \epsfbox{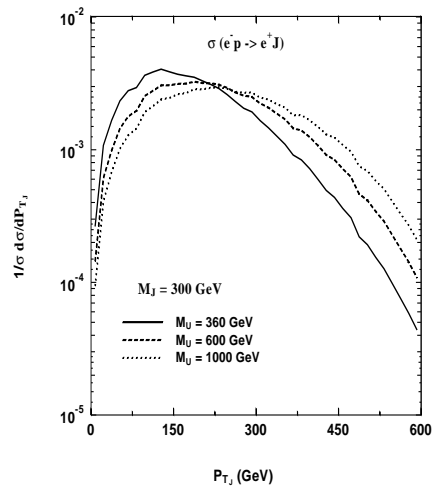}

\end{center}

\vskip -1 true cm
\caption[]{Normalized $P_{TJ}$ distribution with $M_J = 300$ GeV for various
$M_U$.}
\label{fig:5}

\end{figure}

On the other hand, in DIS process a $J$ quark production dominates the top quark 
production in the standard model, since we are using the aforementioned upper 
bound for the 3-3-1 quark mixing parameter which is greater than the analogous 
$V_{\rm CKM}$ one. In the same way as we comment in the Sec. \ref{intro} that we do 
not know as the 3-3-1 Higgs scalars contribute to the muonium to antimuonium 
conversion, we also do not can predict how would be its contribution to the quark
production in DIS. 
\begin{figure}[h] 
\begin{center} 
\epsfxsize=3.8in 
\epsfysize=9 true cm 
\vskip -1.6 true cm \hskip 2 true cm
\ \epsfbox{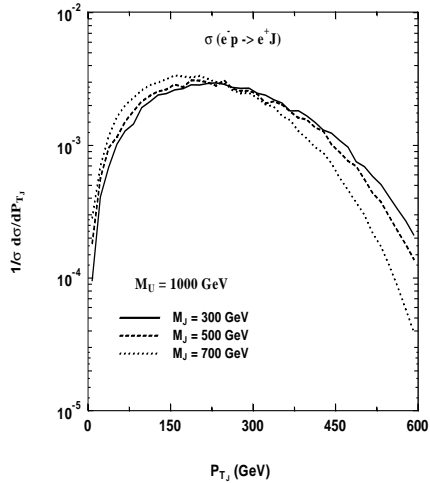} 
\end{center} 
\vskip -1.0 true cm
\caption[]{Normalized $P_{TJ}$ distribution with $M_U = 1000$ GeV for various
$M_J$.} 
\label{fig:6} 
\end{figure} 
For sake of simplicity we is assuming here that the gauge
boson contributions dominate in this process.\par 
\begin{figure}[h]
\begin{center}
 
\epsfxsize=3.8in
 
\epsfysize=9 true cm
 
\vskip -1.7 true cm \hskip 1 true cm 
\ \epsfbox{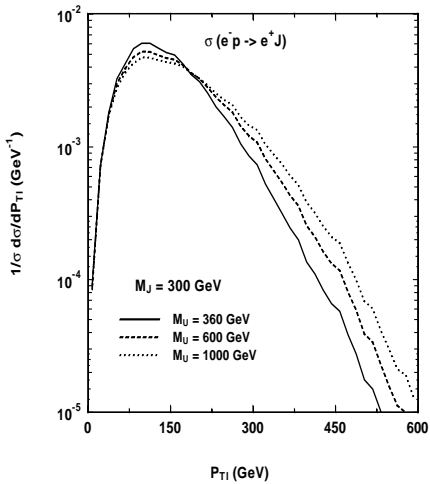}

\end{center}

\vskip -1 true cm
\caption[]{Normalized $P_{Tl}$ distribution of the more energetic secondary
lepton with $M_J = 300$ GeV for various values of $M_U$.}
 
\label{fig:7}

\end{figure}

The Fig. \ref{fig:2} shows the total cross section for heavy quark production as a 
function of quark mass ($M_J$) for some values of double charged bilepton mass. 
For CERN LEPII-LHC center of mass energies (1.8 TeV) and an integrated 
luminosity of ${\cal L}= 6$ fb$^{-1}$yr$^{-1}$, we obtain for $M_J = 300$ GeV 
between 70 and 750 events/year for $M_U$ running from 1200 GeV to 360 GeV. For $M_J = 
600$ GeV between 20 and 120 events/year are expected for the same
range of $M_U$. Would be notice that the limits of these ranges of events
correspond to the particular case when the leptonic mixing parameter is
diagonal (see comment in Sec. \ref{intro}). We observed that the cross
section decreases as  $M_J$ and $M_U$ increase.\par
\begin{figure}[h]
\begin{center} 
\epsfxsize=3.8in 
\epsfysize=8.0 true cm 
\vskip -1.6 true cm 
\hskip 1 true cm
\ \epsfbox{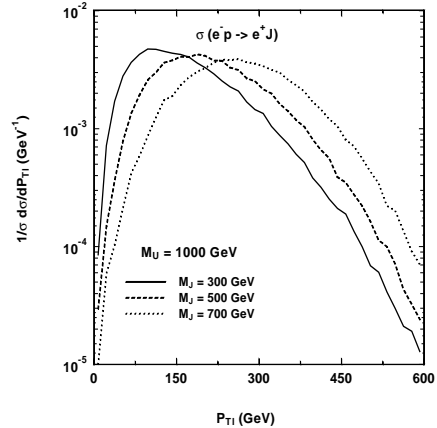} 
\end{center} 
\vskip -1 true cm
\caption[]{Normalized $P_{Tl}$ distribution of the more energetic secondary
lepton with $M_U = 1000$ GeV for various values $M_J$.}  
\label{fig:8} 
\end{figure}

The primary lepton normalized angular distribution can be seen in Figs. \ref{fig:3} 
and \ref{fig:4}. In the former one we fixed $M_J$ = 300 GeV and studied the 
distributions for some values of $M_U$. In the latter same as done for $M_J$ = 500 
GeV. 
\begin{figure}[h] 
\begin{center} 
\epsfxsize=3.8in 
\epsfysize=9 true cm 
\vskip -1.6 true cm 
\hskip 1 true cm
\ \epsfbox{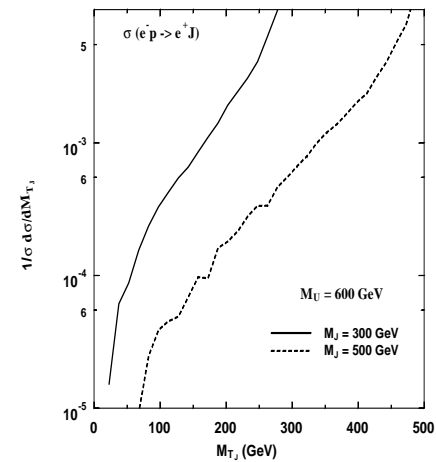} 
\end{center} 
\vskip -1 true cm
\caption[]{Normalized cluster transverse mass spectrum of the exotic heavy
quark.}  
\label{fig:9} 
\end{figure}
We observed  that these distributions dominate the backward region, the increase 
of quark  mass difficulties to distinguish the plots for different bilepton 
masses.\par 
The transverse quark momentum distribution can be seen in Fig. \ref{fig:5} for fixed  
$M_J$ = 300 GeV and for some values of $M_U$. We observed that the maximum value  is 
for $p_T \simeq$ 150 GeV, for an expressive range of bilepton masses. The same 
behavior is observed to other values of $M_J$ which was not showed. In Fig. 
\ref{fig:6} we fixed $M_U$ = 1000 GeV and we can see that the shapes of the curves 
are  very similar among them.\par 
We continue our analysis, by using the same values as the Figs. \ref{fig:5} and 
\ref{fig:6}, now studying the transverse momentum of the most energetic secondary 
lepton produced in the  $J_1$ quark decay (see Fig. \ref{fig:1}). As the Fig. \ref{fig:5}, 
the Fig. \ref{fig:7} does not allow to distinguish the plots. On the other hand, Fig. 
\ref{fig:8} shows distributions whose maxima values in $p_T$ are shifted to right as 
$M_U$ increases. For other values of $M_J$ the plots present similar behavior.\par

Finally we present in Fig. \ref{fig:9} the transverse quark mass which gives 
information  about the recoiling $u$ quark jet and has a sharp peak at the exotic 
quark mass.

\section{Conclusions}

We have investigated the production and decay of one of the exotic quark 
predicted by 3-3-1 electroweak model at CERN LEPII-LHC center of mass energy. 
Employing a Monte Carlo simulation we studied the process $e^- p \longrightarrow 
e^+ X l^- l^-$ depicted in Fig. \ref{fig:1}. We analyzed the total cross section as a 
function of exotic quark mass and several kinematical distributions of the final 
state particles in order to identify the masses of the new particles of the 
model. We stress that we have an exclusive semileptonic channel of heavy quark 
decay in the model, which is free of the standard model background and can gives 
a reasonable rate of events in the CERN LEPII-LHC center of mass energy. In the 
range of typical values of mass parameters we have between 20 and 750 
events/year. \par
We can see that the $p_T$ distribution of the more energetic secondary lepton is 
more sensitive to $M_J$ variation (see Fig. \ref{fig:8}). It shows clearly a distinction 
between the exotic quark masses for a fixed double charged bilepton mass. One 
should also notice that the Fig. \ref{fig:8} is a transverse momentum distribution of the 
most energetic secondary  lepton which is directly observed in the 
experiment.\par
 We conclude so that the 3-3-1 model is a promising option for searching new 
physics in the next generation of accelerators and the deep inelastic scaterring 
could gives spectacular signatures.

\label{sec3}

\acknowledgements
We would like to thank Professors J. A. Martins Sim\~oes and V. Pleitez for 
helpfully comments and suggestions. One of us (M. D. T.) would like to thank the 
Funda\c c\~ao de Amparo \`a Pesquisa no Estado do Rio de Janeiro for full 
financial support (contract No. E-26/150.338/97).

\end{document}